\documentclass[fleqn,usenatbib]{mnras}

\usepackage{newtxtext,newtxmath}

\usepackage[T1]{fontenc}

\DeclareRobustCommand{\VAN}[3]{#2}
\let\VANthebibliography\thebibliography
\def\thebibliography{\DeclareRobustCommand{\VAN}[3]{##3}\VANthebibliography}

\usepackage{graphicx}	
\usepackage{amsmath}	
\usepackage{supertabular}
\defcitealias{Weltevrede2006}{W06}
\defcitealias{Malofeev2018}{M18}

\title[Drift periods from summed power spectra]{Pushchino multibeams pulsar search - III. Drift periods of pulsars from summed power spectra method}

\author[T. V. Smirnova et al.]{
T. V. Smirnova,$^{1}$
S. A. Tyul'bashev,$^{1}$\thanks{E-mail: serg@prao.ru (SAT)}
M. A. Kitaeva,$^{1}$
V. M. Malofeev $^{1}$
\\
$^{1}$ P.N. Lebedev Physical Institute of the Russian Academy of Sciences, Astro Space Center, Pushchino Radio Astronomy Observatory,\\
Radiotelescopnaya 1a, Moscow reg., Pushchino, 142290, Russia \\
}

\date{January 24, 2020}

\pubyear{ , 2023}

\begin{document}
\maketitle

\begin{abstract}
The drift periods $P_2$ and $P_3$ were searched for using the summed power spectra of 41 pulsars observed at declinations from $-9\degr$ to $+42\degr$. The power spectra of pulses with a given period, pulse width and drift behavior have been simulated, the applicability of such a method for estimating drift parameters is shown. For most pulsars, the distribution of harmonic amplitudes in the power spectra corresponds to the expected distribution for these pulsars without drift. At the same time, it was found that for a number of sources, the summed power spectra accumulated over a long period of time give the same drift parameters as those determined by other methods. For 11 pulsars we have defined or redefined the drift period $P_2$. For 8 sources the drift period $P_3$ has been determined or redefined. The drift direction of subpulses was redefined for them.
\end{abstract}





\section{Introduction}

Individual pulsar pulses consist of one or more subpulses appearing at certain longitudes in the average profile window.  Practically since the discovery of pulsars, for some of them, a change in the position of the subpulses was detected over time inside the window of the average profile, that is, the drift of subpulses (\citeauthor{Drake1968}, \citeyear{Drake1968}). The drift manifests itself in a regular shift of the subpulse phase when observing a sequence of pulses. The subpulses form ``drift bands'', the horizontal separation between them corresponds to the period $P_2$. This period has a range from several to hundreds of milliseconds.  The position of the subpulse in the window of the average profile is repeated after a period of time that is expressed in pulsar periods $P_1$. This interval, which is generally not an integer number, is called a period $P_3$. Obviously, to get estimates of periods $P_2$ and $P_3$ it is necessary to register individual pulses over a long time interval, which is a problem, since the observed intensity of individual subpulses is usually low. Subpulse drifting is observed in more than a third of known pulsars (\citeauthor{Weltevrede2006} (\citeyear{Weltevrede2006}) here and further on \citetalias{Weltevrede2006}, (\citeauthor{Song2023} (\citeyear{Song2023})). The largest number of $P_2$ and $P_3$ estimations based on observations of individual pulses were done in the paper \citetalias{Weltevrede2006}. Of the 187 pulsars observed in Westerbork at the frequency of 1,400 MHz, 170 pulsars have different kinds of subpulse modulation, and 68 pulsars, that is, about one third of them, show drift behavior. In the paper \citetalias{Weltevrede2006} according to the drift behavior, pulsars are divided into 3 classes: Coh, Dif and Dif* depending on the width of the spectral details in 2DFS spectra. If the detail in the spectrum is narrow (less than 0.05 of cycle for a period), then the drift is coherent (Coh). Pulsars showing wide diffuse details are divided into Dif* (the details are close to the beginning or the end of the spectrum) and Dif (diffuse detail are far from the edges of the spectrum).

Presumably, the drift behavior of pulsar subpulses is an internal feature of their radiation mechanism, therefore, its study is an important physical problem. The estimation of  $P_2$ and $P_3$ periods can be influenced by different factors. One of them is nulling, that is, the skipping of successive pulses. In case of nulling, the radiation mechanism is ``switched off'',  and the drift velocity can become less and then return to the previous value in some cases (\citeauthor{Janssen2004}, \citeyear{Janssen2004}). Another effect observed for a number of pulsars is switching a mode (shape) of the integral profile. When switching modes, the drift velocity may change (\citeauthor{Huguenin1970}, \citeyear{Huguenin1970}; \citeauthor{Izvekova1982}, \citeyear{Izvekova1982}). Up to now, several models are known to explain the drift behavior of pulsars.  The most widely used one is sparking gap model  (\citeauthor{Ruderman1975}, \citeyear{Ruderman1975}), developed in the works of many authors (\citeauthor{Gil2003}, \citeyear{Gil2003}; \citeauthor{Qiao2004}, \citeyear{Qiao2004}). According to this model, the drift behavior can be explained by the presence of rotating discharges which circulate around the magnetic axis under the action of ${\bf E} \times {\bf B}$ drift.  In such a model, the arrival times of the pulses should shift in the window of the average profile, and pulsars should show certain periods of circulation. In a number of works, the periodicity in the modulation of the amplitude of subpulses has been interpreted as caused by the regular rotation of plasma ``sub-beams'' around the magnetic axis (carousel model), and the time of such circulation was determined (\citeauthor{Deshpande2001}, \citeyear{Deshpande2001}; \citeauthor{Gupta2004}, \citeyear{Gupta2004}) for several pulsars.

As a rule, to measure drift periods $P_2$ and $P_3$ the spectra from the amplitude time variations of  the subpulses at different longitudes of the average profile are used. Obviously, getting estimates of $P_2$ and $P_3$ is technically challenging task, since this requires observations of individual pulses with a good signal-to-noise ratio ($S/N$) and, therefore, a high fluctuation sensitivity of the radio telescope is needed. Analysis of integral two-dimensional (by time and longitude) fluctuation spectra (2DFS) allowed to study the drift behavior of a large number of pulsars (\citeauthor{Edwards2002}, \citeyear{Edwards2002}; \citeauthor{Edwards2003}, \citeyear{Edwards2003}), \citepalias{Weltevrede2006}. 

In the paper (\citeauthor{Malofeev2018} (\citeyear{Malofeev2018}) (here and further on \citetalias{Malofeev2018}) another method of $P_2$ and $P_3$ evaluation was used. When searching for pulsars at the frequency 111 MHz in the monitoring data from Large Phased Array (LPA) of the Lebedev Physics Institute (LPI) power spectra from time series of pulsed pulsar emission summed over many days were used to increase sensitivity (\citeauthor{Tyulbashev2017}, \citeyear{Tyulbashev2017}). It turned out that there may be characteristic features in the power spectra the observed frequency of which is associated with periods of subpulse drift $P_2$ and $P_3$.  The amplitude of successive harmonics, the first of which is determined by the pulsar period $P_1$,  can decrease sometimes to the level of the noise, then it begins to grow and decrease again. There are often several such ``humps''. In the paper \citetalias{Malofeev2018}, such modulation was associated with the drift period $P_2$. The second feature in the power spectra is harmonics-satellites, when an additional harmonic appears near to the main harmonics. The spacing between the satellite harmonic and the main harmonic was related to the drift period $P_3$. Using the summed power spectra of 27 strongest pulsars in the declination range $+21\degr < \delta < +42\degr$, in the work of \citetalias{Malofeev2018} were obtained the estimations of $P_2$ for 26 pulsars, and $P_3$ for 13 pulsars. 

In this paper, the search for drift periods $P_2$ and  $P_3$ is carried out from the summed power spectra of 41 pulsars observed on declinations from $-9\degr$ to $+42\degr$. Summation of power spectra makes it possible to significantly increase S/N and identify weak details in the power spectrum for a weak pulsars. As will be shown in our work, the analysis of the summed power spectrum makes it possible to determine the most frequently implemented drift parameters, including the drift direction, most characteristic of a pulsar. It is known that for a number of pulsars, the direction and drift parameters for sub-pulses can change over time \citetalias{Weltevrede2006} (\citeauthor{Edwards2003}, \citeyear{Edwards2003}),  which significantly complicates the overall picture. Our method allows, unlike other methods, to determine the average parameters of pulsar drift over a long- time span. The simulation of the drift behavior of pulses from selected pulsars and the calculation of power spectra from them was done, taking into account: the time interval of observations used in one session, the period of pulsar, width of subpulses and the modulation of the flux density from pulse to pulse. The simulated power spectra were compared with the power spectra obtained as a result of summing the power spectra over many observation sessions, and $P_2$ and $P_3$ were determined. We also revised the early estimates of $P_2$ and $P_3$  from the paper \citetalias{Malofeev2018}.

\section{Monitoring program of observations at the LPA LPI}

LPA LPI antenna is an array built on wave dipoles. The signal coming from the dipole lines comes to the first level of the amplifiers and multiplies into four outputs. This makes it possible to create four independent radio telescopes based on one antenna array. Of the available four outputs, one output is used for maintenance and quality control of the antenna, and two outputs are used for scientific purposes. That is, two independent radio telescopes are currently operating on the basis of one antenna. One of the radio telescopes is operating under a long-term monitoring program. This radio telescope has 128 beams located in the meridian plane and overlapping declinations from $-9\degr$ to $+55\degr$. For 96 beams, digital receivers were made in 2014 along with the completion of the LPA upgrade, and the remaining beams were connected to the receivers in test mode less than a year ago. LPA is a meridian type telescope. The observation time for one day (one session) is limited by the time the source passes through the meridian and is approximately 3.5 minutes at half the power of the radiation pattern. Since monitoring is daily and round-the-clock, it was possible to accumulate about five days of data for each point in the sky for an interval of 5.5 years (2057 power spectra).

The same monitoring data are used for research in two scientific areas. Firstly, daily observations of several thousand radio sources scintillating on the interplanetary plasma provide a fundamental opportunity to predict ``Space Weather'' using an antenna array (\citeauthor{Shishov2016}, \citeyear{Shishov2016}). Secondly, observations are used to search for pulsars and rotating radio transients (\citeauthor{Tyulbashev2016}, \citeyear{Tyulbashev2016}; \citeauthor{Tyulbashev2018}, \citeyear{Tyulbashev2018}) within the framework of the project Pushchino Multibeams Pulsar Search (PUMPS, (\citeauthor{Tyulbashev2022} (\citeyear{Tyulbashev2022})).Two modes are used simultaneously for the monitoring program. In the first mode, the recording is in the band 2.5 MHz in six-channel mode with channel width 415 kHz and the sampling frequency 10 Hz. In the second mode, data are recorded in thirty-two channels in the 2.5 MHz band with a channel width of 78 kHz and the sampling frequency of 80.0461 Hz. The time interval of each individual record is: $T = 204.682$~s and, accordingly, the length of the record is 16,384 points. The main task for this mode is to search for pulsars and transients of the RRATs type. In this paper, we use the data obtained in the second mode. For each record we have a sequence of pulses with a pulsar period $P{_1}$. For each such record, we calculated the power spectrum using Fast Fourier Transform (FFT). As a result, we had 8192 points of the power spectrum for each session, the first (main) harmonic of which has the position $T/P_1$. The time resolution was 12.4928 ms.

To increase the sensitivity, the search for pulsars is carried out using power spectra summed up over all monitoring data after excluding records corrupted by interference and the ionosphere. It turned out that for some strong pulsars, more than 100 harmonics can be observed in a summed power spectrum \citepalias{Malofeev2018}. 

\section{Criterion for the selection of power spectra} 

In the power spectrum, the first ``hump'', associated with the modulation of the main harmonics of the power spectrum and possibly caused by drift, can fall at the end of a power spectrum. 
For example,  the position of the main harmonic in the power spectrum is defined as $T/P_1$, and for a pulsar with a period $P_1 = 100$~ms it will be at 2,047 point. In total we have 8,192 points in the power spectrum. Therefore the second harmonic will be at 4,094 point and only 4 harmonics will be included in the full power spectrum. For a confident determination of $P_2$, it is necessary to have more than 5 harmonics to see modulation. For pulsars with a period less than 100 ms, the search for $P_2$ in the monitoring data will be impossible.  The criterion for selecting pulsars to search for drift is using pulsars with $P_1 > 200$~ms, for which more than 8 harmonics can be observed in the power spectrum for possible detection of harmonic modulation. In this case we can detect at least two ``humps''. A total of 41 pulsars with $S/N > 10$ were chosen for the first harmonic in the power spectrum observed on declinations from $-9\degr$ to $+42\degr$ (the list of pulsars see in the Appendix~\ref{Appendix A}).

To obtain the summed power spectrum, high-quality data was selected, and the rest was discarded. To do this, we calibrated the raw data in the frequency channels using a calibration signal, which is recorded 6 times a day in all beams. After that, the standard deviations were evaluated and only those power spectra that improve the S/N were selected for summation. All these procedures are reflected in detail in the paper by (\citeauthor{Tyulbashev2020} (\citeyear{Tyulbashev2020}).

\section{Data analysis}

To obtain estimates of pulsar drift periods, we used the summed power spectra after dispersion removal and gain equalization in frequency channels. Before calculating the power spectra, the data were cleaned from interference and the noise level was estimated. A calibration signal of a known temperature is recorded every four hours. This allows to calibrate data for different days in the same units. The average value of noise sigma  ($\sigma_N$) using the calibrated signal was evaluated for each source independently  and, if in this record $\sigma_N$ exceeded the specified noise level for this direction (see details in the paper (\citeauthor{Tyulbashev2020} (\citeyear{Tyulbashev2020}), then it was not used for further processing. For each selected record, the power spectrum was calculated by the Fast Fourier Transform method (FFT). Then the power spectra were summed up for all the observation sessions. As a result, we had $N/2$ points of the average power spectrum with a resolution of $\Delta f = 0.004886$~Hz. The total number of power spectra summed over ranged between 700 to 2,000.

\subsection{Pulses drift model} 
\label{subsecP}

For analysis of the summed up pulsar power spectra, we used the following model. A sequence of pulses with a pulsar period $P_1$ was formed on the time interval $T$. It was assumed that the pulses have a Gaussian shape, their amplitude and half-width at the level of $1/2$ of the amplitude ($W_{0.5}$) were set as a parameters. All pulsars studied by us have small dispersion measures: less than 51~pc/cm$^3$, so pulse broadening due to scattering in interstellar plasma at a frequency of 111 MHz is less than our time resolution (12.5 ms) and therefore does not affect the shape of the pulse. The pulse amplitude distribution was also a parameter. We considered 2 options: 1) constant pulse amplitude with the value A, 2) randomly distributed pulse amplitudes, $a_i$, according to the equation: $a_i = (rand-0.5)\times 2A \times 0.01smod+A$, where $smod$ is the percentage of modulation (from 0 to 100\%) and $rand$ is a random number from 0 to 1. We used here the random number generator. The mean value of $A$ used for simulation is 10 (in arbitrary units). $P_2$ can be either positive or negative. A negative value of $P_2$ means that the subpulses appear earlier in successive pulses (subpulses are shifted to the left, negative drifting) or to the right ($P_2$ is positive, positive drifting). Signs of $P_2$ therefore correspond to the drift direction, such that a positive sign corresponds to positive drifting. One or two subpulses with a spacing between them equal to $P_2$ ms were set in each period having equal half-width and amplitude. The phase shift of the pulses during one pulsar period was determined by the ratio $P_2/P_3$, that is, after every $P_3$ pulsar periods the phase of the next pulse returns to the initial position before the shift. The direction of drift in the simulation was one of the parameters. 
In the presence of one subpulse, 2 options were considered: 1) in the absence of the drift, it was assumed that $P_2$ is equal to zero, and it does not shift in time; 2) the subpulse shifts with a velocity $P_2/P_3$. Based on the received sequence of pulses with or without drift, the power spectrum was calculated using FFT, and a comparison was made with the resulting summed up power spectrum. In the power spectrum $P_2$ will manifest itself as a periodic modulation of the main harmonics of the power spectrum with a period $k_{p_2} = T / P_2$. $P_3$ will appear as additional harmonic satellites at a spacing $r = T / (P_3 \times P_1)$ from the main harmonics ($r$ is the spacing expressed in points, and  $k_{p_2}$  is the number of points in the power spectrum of the first ``hump'' of modulation).

\begin{figure}
	\includegraphics[width=\columnwidth]{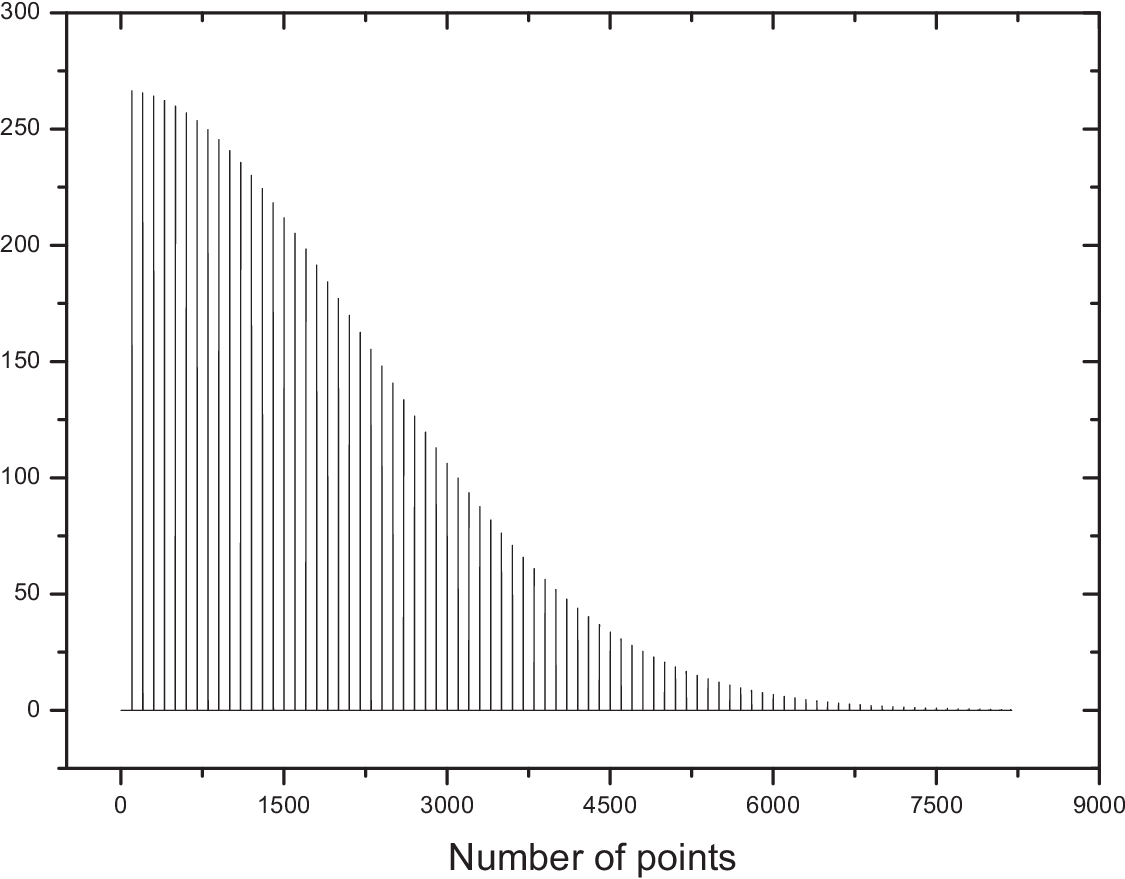}
	\caption{Power spectrum for the model array with $P_1 = 2.04682$~s, $W_{0.5}  = 12.5$~ms. The horizontal axis shows the points  number in the power spectrum ($\Delta f = 0.004886$~Hz), the vertical axis shows the amplitude of harmonics in arbitrary units}
    \label{fig:fig1}
\end{figure}

\begin{figure}
	\includegraphics[width=\columnwidth]{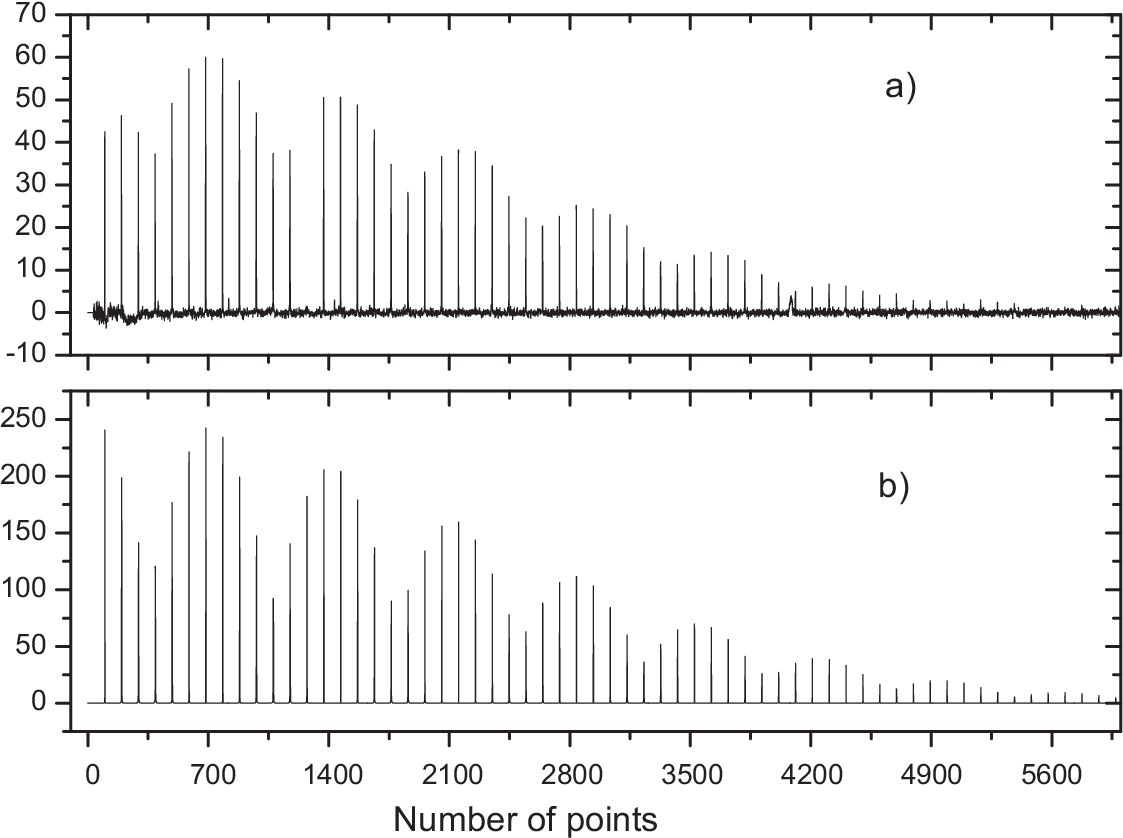}
	\caption{ a) The resulting summed up power spectrum, b) power spectrum model for pulsar  J0928+3037. Period $P_1 = 2.0915$~s, $W_{0.5}  = 12.5$~ms,  $P_2 = 0$~ms. The designation of the axes is the same as in Fig.\ref{fig:fig1}. 
 }
    \label{fig:fig2}
\end{figure}

Since we have a finite length of the data record $T$  and, in most cases, the ratio $T/P_1$ is not an integer number, there is an additional modulation of the harmonic amplitudes  of the power spectrum. The harmonic amplitude increases as $k \times (T / P_1)$ approaches to an integer number ($k$ is harmonic number). Let's explain it with an example. If $T / P_1$  were an integer number, then the amplitude of harmonics in the power spectrum would decrease in accordance with the Gaussian envelope, the half-width of which is proportional to the inverse width of the initial pulses. Fig.\ref{fig:fig1} shows an example of such a model power spectrum for a sequence of pulses with $P_1 = 2.0468$~s, multiple of $T$,  and pulses half-width $W_{0.5} = 12.5$~ms. For the point number in the power spectrum $n \ge 6,000$ the amplitude of harmonics drops by more than 250 times, and it is possible to detect modulation associated with drift only for very strong pulsars. Fig.\ref{fig:fig2}b shows an example of a model power spectrum from a sequence of pulses with the period $P_1 = 2.0915$~s without drift (the period corresponds to PSR J0928+3037) and $W_{0.5}  = 12.5$~ms. In this case $T / P_1 = 97.845$ and 6th, 7th, 8th harmonics, and also multiples of them, have the values of the position of harmonics closest to an integer numbers $kT/P_1$: 587.07, 684,92,782.76, which leads to their local maxima. Accordingly, periodic modulation of harmonic amplitudes appears, which is not associated with pulses drift. Fig.\ref{fig:fig2}a shows the summed up power spectrum obtained from observations, on which the same harmonic modulation is visible. The absence of the thirteenth harmonic is caused by interference and it was deleted. For model pulses, the half-width $W_{0.5}  = 12.5$~ms, is mainly used, which adequately characterizes the real width of individual pulses (25 ms total width at the level of $1/2$ of maximum) at our sampling time 12.5 ms. The pulse width does not affect the simulation results. However we should point that the amplitude of harmonics will be modified depending on width, which should have some effect on the power spectral properties. The amplitude of the harmonics will go as $\sim  W_{0.5} ^2$ and the half-width of the harmonic distribution in the power spectrum will be proportional to $1/W_{0.5}$. This was verified by simulating pulses with different $W_{0.5}$. Even the width shows variation between different data and pulsars, the summed up power spectrum reflects an average width of the pulses in the observation range.

\begin{figure}
	\includegraphics[width=\columnwidth]{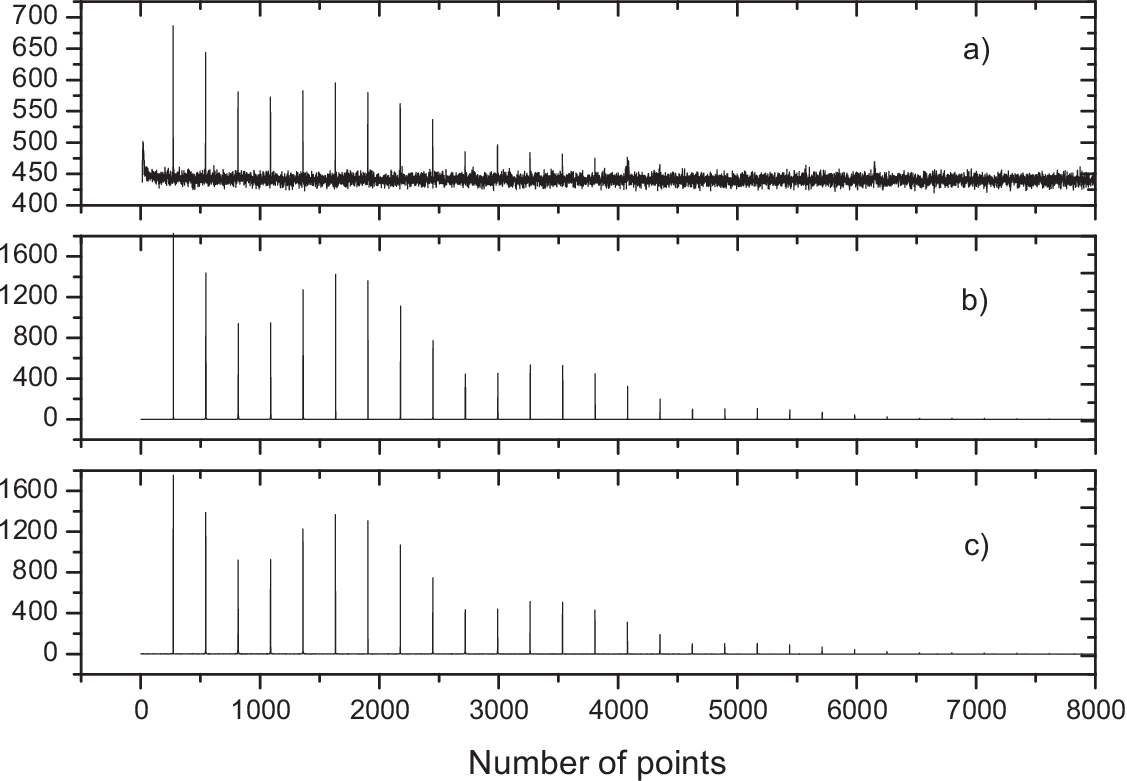}
	\caption{Power spectrum for pulsar J1823+0550 ($P_1 = 0.75291$~s): a) the resulting summed up power spectrum; b) the model without pulse amplitude modulation; c) model with 50\% pulses modulation on amplitude, $P_2 = 0$. The designation of the axes is the same as in Fig.\ref{fig:fig1}}
    \label{fig:fig3}
\end{figure}

Modeling has shown that random variations in the amplitudes of the initial pulses have little effect on the distribution of the amplitudes of the power spectrum harmonics (see Fig.\ref{fig:fig3}). It can be seen from the figure that the model arrays describe the observations quite well.

\subsection{The occurrence of drift in power spectra}

\begin{figure}
	\includegraphics[width=\columnwidth]{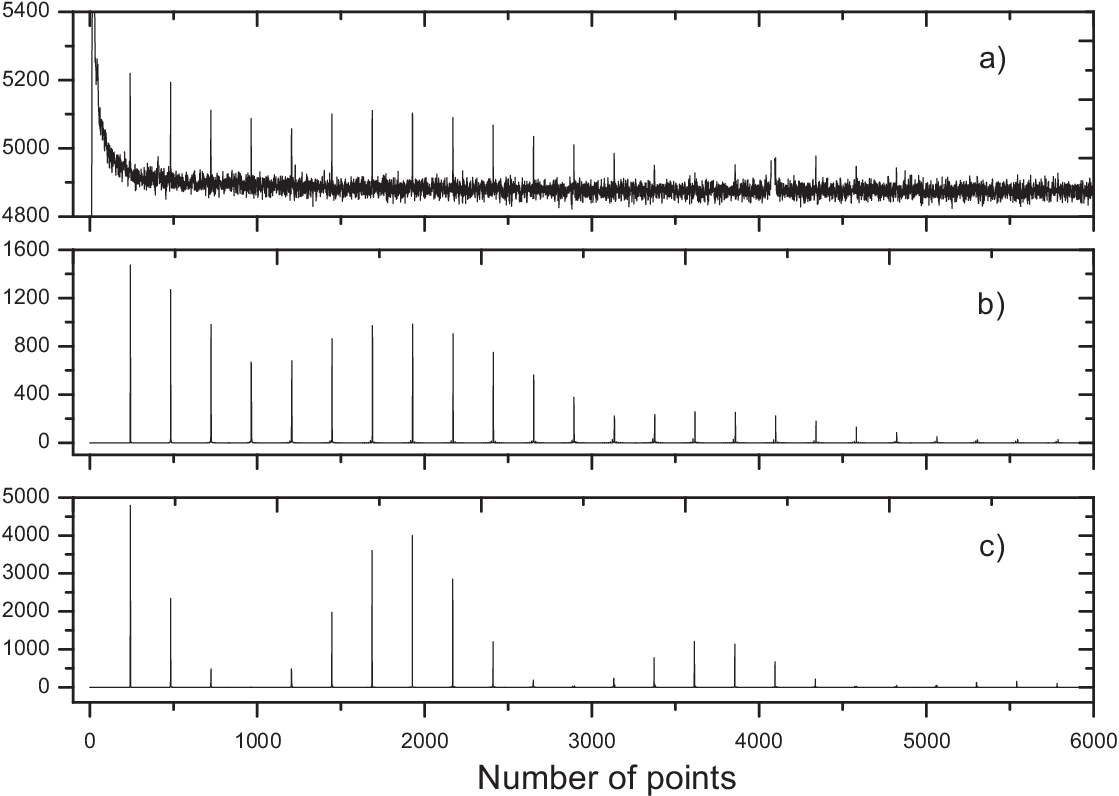}
	\caption{a) The resulting summed up power spectrum PSR J1313+0931 ($P_1 = 0.8489$~s, $P_2$ = 0 ms); b) the drift-free power spectrum model ($P_1 = 0.8489$~s, $P_2=0$~ms); c) added drift with parameters  $P_2 = 108$~ms, $P_3 = 122P_1$. The designation of the axes is the same as in Fig.\ref{fig:fig1}}
	\label{fig:fig4}
\end{figure}

Fig.\ref{fig:fig4} shows an example of the summed up power spectrum (Fig.\ref{fig:fig4}a) and model power spectra for J1313+0931: without drift (Fig.\ref{fig:fig4}b) and with drift (Fig.\ref{fig:fig4}c). The value of $P_2$ corresponds to the harmonic with the number $T /P_2 = 1,895$, at which the first maximum in the distribution of harmonics is observed.  Since there are no additional harmonic satellites in the observed power spectrum next to the main ones, it means that $r$ (distance to the main harmonic in points) $\le 2$ (we just won't see the extra harmonic), accordingly $P_3 > T / (2 P_1)$,  this is true for this model ($T / (P_3 P_1) = 1.98$), but there is a deep modulation of harmonics. It can be seen from the simulation that the drift-free model agrees much better with the observations. Let us note that in the presence of a two-component structure of the average profile, even without pulse drift, a deep modulation of harmonics associated with the distance between subpulses will also be observed in the power spectrum. It is obvious that the observed ``humps'' in the power spectra can be associated with the real drift of subpulses, but this is not a sufficient condition for a conclusion about drift. Modulation of the amplitudes of the power spectrum harmonics   can also occur due to the fact that $T / P_1$ is not an integer number, as shown above (see Fig.\ref{fig:fig2}). When defining parameters $P_2$ and $P_3$, it is necessary to take into account these effects. As the simulation shows, the drift manifests itself in a power spectrum as the appearance of additional harmonics to the left or right of the main harmonics, depending on the direction of drift. Sometimes additional harmonic satellites appear both to the left and to the right of the main harmonic, but their amplitudes can differ significantly. Period sign $P_2$ is determined by a harmonic with a larger amplitude.

\begin{figure}
	\includegraphics[width=\columnwidth]{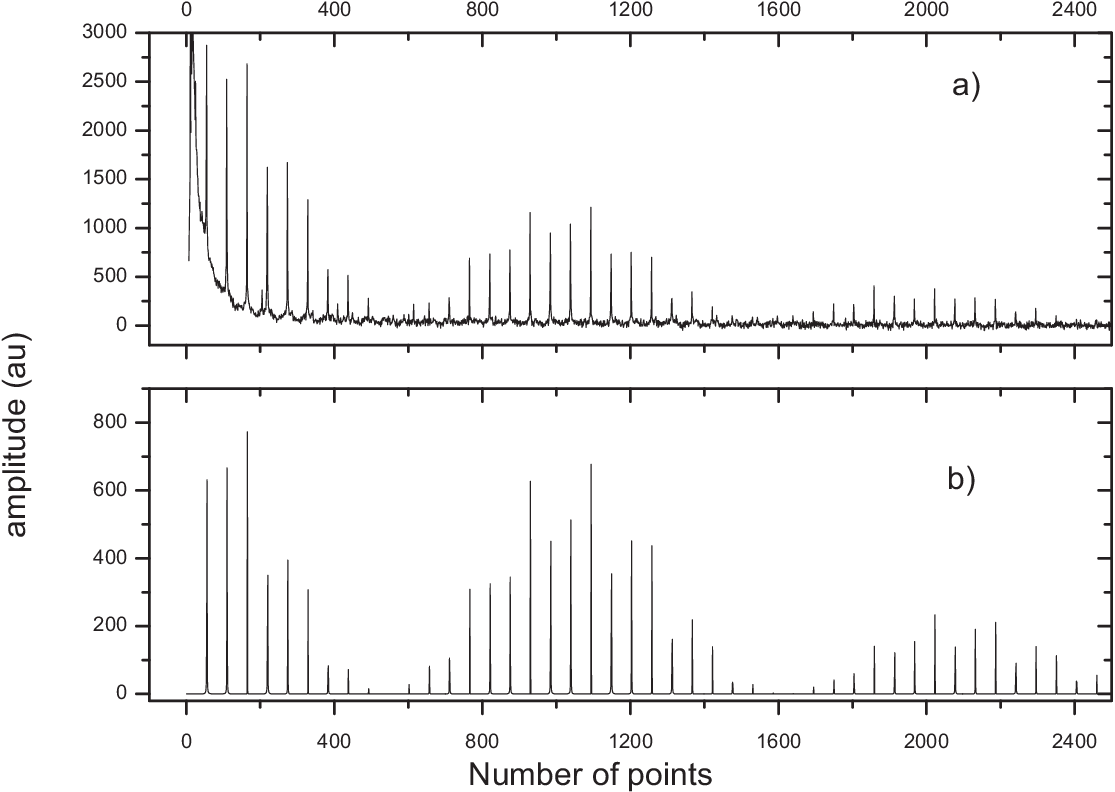}
	\caption{a)  Power spectrum for pulsar J0528+2200 after summing up 600 sessions; b) the model power spectrum without drift corresponding to the following pulsar parameters: $P_2 = 190$~ms, $W_{0.5} = 22$~ms. The designation of the axes is the same as in Fig.\ref{fig:fig1}}
	\label{fig:fig5}
\end{figure}

\begin{figure}
	\includegraphics[width=\columnwidth]{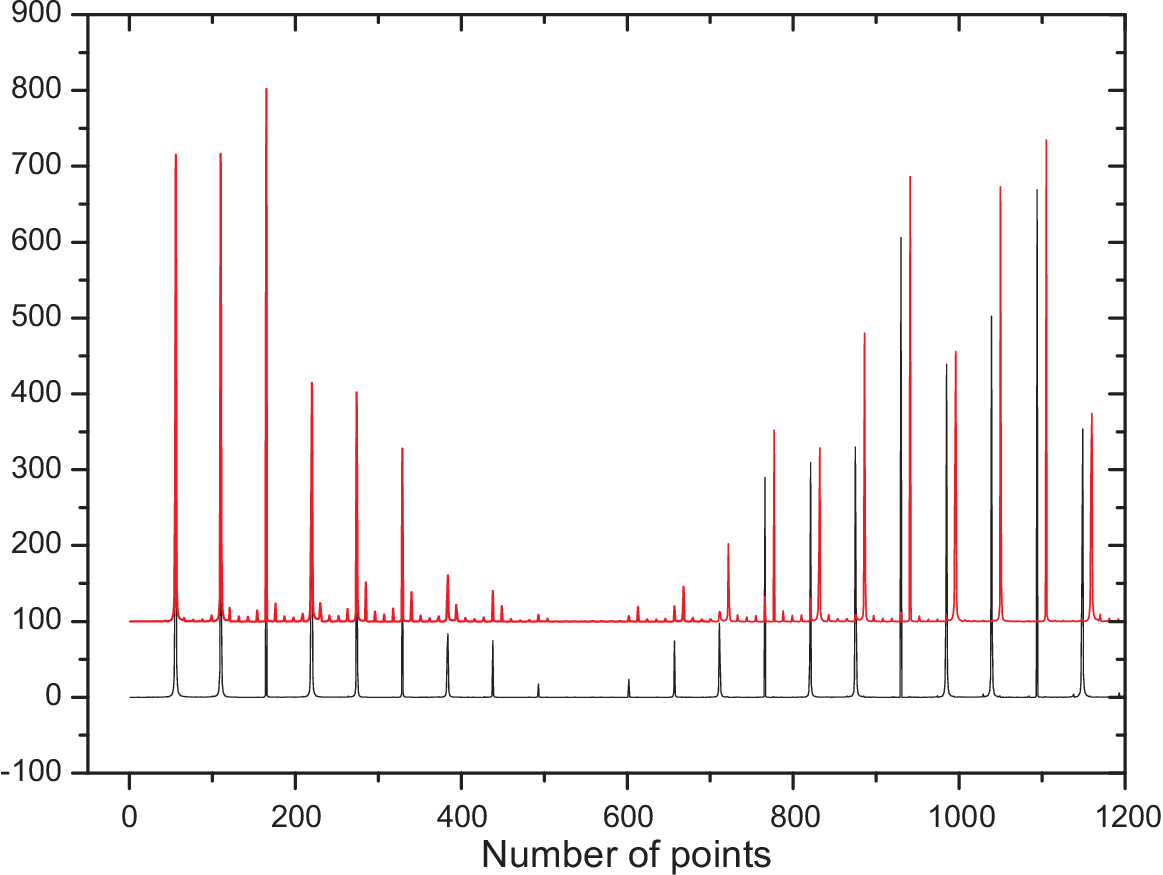}
	\caption{Model power spectra for J0528+2200: without drift (black line); with drift: $P_2 = -190$~ms, $P_3 = 4.9P_1$; $W_{0.5} = 22$~ms, the power spectrum is shifted up along the y-axis for clarity. The designation of the axes is the same as in Fig.\ref{fig:fig1}}
	\label{fig:fig6}
\end{figure}

Fig.\ref{fig:fig5}a shows an example of the accumulated power spectrum for 600 sessions for PSR    J0528+2200 (B0525+21). Additional harmonic satellites are not visible in the resulting power spectrum. The main harmonics do not have displacements, in comparison with their expected position, and, therefore, there is no drift. The modulation of the main harmonics of the power spectrum with clearly expressed two maxima is clearly visible. This pulsar has a 2-component structure of a average profile with a distance between the components at a frequency of 111 MHz equal to190 ms and a half-width components equal to 22 ms (\citeauthor{Smirnova2009}, \citeyear{Smirnova2009}).  The simulation calculation of the power spectrum for the 2-component structure with a distance of 190 ms without drift (Fig.\ref{fig:fig5}b) and with drift to the beginning of the profile (Fig.\ref{fig:fig6}) shows a visible good agreement with the summed up power spectrum for both models. However, the inclusion of drift leads to a rapid decrease in the amplitude of the main harmonics and to the dominance of harmonics shifted from the main ones by $r = T / (P_1 P_3)$ points. This can be seen in Fig.\ref{fig:fig6}, which shows, shifted along the y-axis for clarity, two models on a smaller scale along the x-axis.

In the model with drift, to the right of the main harmonics associated with the pulsar period $P_1$ weak harmonics appear up to $n = 430$, shifted to the right by $r = 11$~points ($P_3 = 4.9P_1$). In the second ``hump'', only the shifted harmonics remain and the main harmonics are practically not visible. The appearance of displaced harmonics is the main manifestation of drift in the power spectra if there is a regular drift. In the paper \citetalias{Malofeev2018}, $P_2$ and $P_3$ have been determined for this pulsar, they correspond to the parameters used by us for the model with drift (Fig.\ref{fig:fig6}). In \citetalias{Malofeev2018} a smaller accumulation of power spectra was used, and in their Fig.\ref{fig:fig4} there is a weak harmonic to the right of the main one, located at a distance corresponding to $P_3$, and, consequently, the direction of drift is towards the beginning of the profile ($P_2$ is negative). This pulsar, as noted in the paper \citetalias{Weltevrede2006} (Table~2), belongs to the class Dif* and has different drift directions and significantly different values of $P_2$ for two profile components. The model describes well the obtained power spectrum with parameters: $P_2 =-190 \pm 16$~ms  and $P_3 = 4.9 \pm 0.2P_1$ agree well with \citetalias{Weltevrede2006}: $P_2 =-200^{+20}_{-90}$~ms and $P_3 = 3.8 \pm 0.7P_1$ for one of the components of the average profile. In \citetalias{Weltevrede2006} was noted that for the other component, the drift goes in the other direction ($P_2$ is positive),  and the analysis gives a wide component in a two-dimensional spectrum. The model we use assumes a regular drift, which is not observed in reality for this pulsar and therefore in \citetalias{Malofeev2018} only a weak harmonic is visible on the right near the main one, and we do not have it with a large averaging of the power spectra.

\begin{figure}
	\includegraphics[width=\columnwidth]{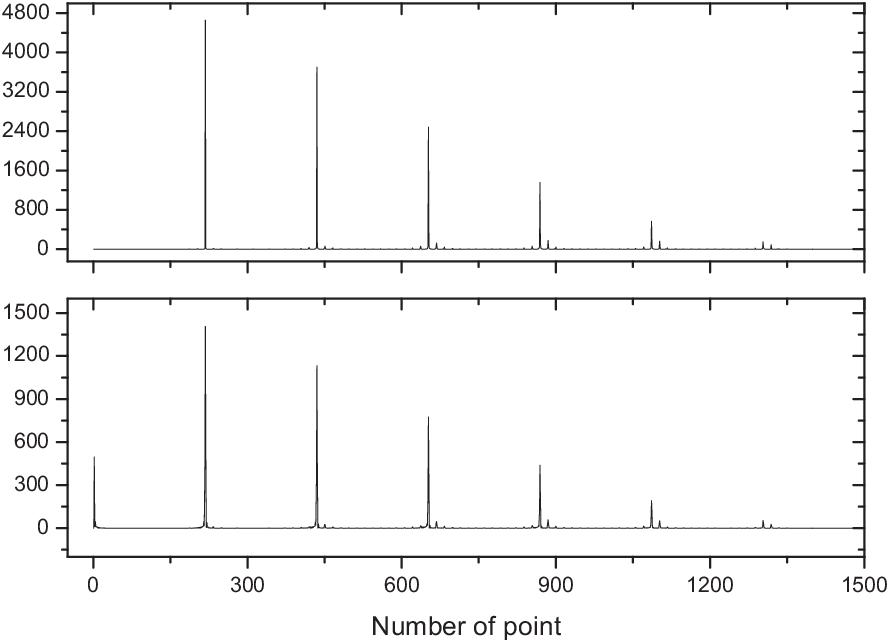}
	\caption{The modeled power spectrum with parameters for PSR J0034-0721. On the top is a part of this power spectrum without nulling. On the bottom is the same for modelling pulses with 45\% nulling duration. The designation of the axes is the same as in Fig.\ref{fig:fig1}}
	\label{fig:fig12}
\end{figure}

The pulsar B0525+21 has a nulling fraction of 25\% (\citeauthor{Wang2007}, \citeyear{Wang2007}). Our model can account for this effect. Simulation of pulses with different nulling durations (up to 50\%) and drift parameters showed that the amplitude distribution of harmonics and the position of satellites in the power spectrum does not change, but the amplitude of all harmonics decreases depending on the nulling length. Sessions with a small fraction of nullings will have the greatest contribution to the power spectrum accumulated over many sessions. The example of spectra for modelling pulses with 45\% nulling duration and without it can be seen from Fig.\ref{fig:fig12}. Off-pulse emission has been reported for this pulsar (\citeauthor{Basu2011} (\citeyear{Basu2011}) but (\citeauthor{Marcote2019} (\citeyear{Marcote2019}) conducted very-high-resolution radio observations of B0525+21 and concluded that the off-pulse emission should be less than 0.4\% of the period-averaged pulsed flux density.  In any case, adding a constant level of radiation outside the pulses will not affect the power spectrum.

\begin{figure}
	\includegraphics[width=\columnwidth]{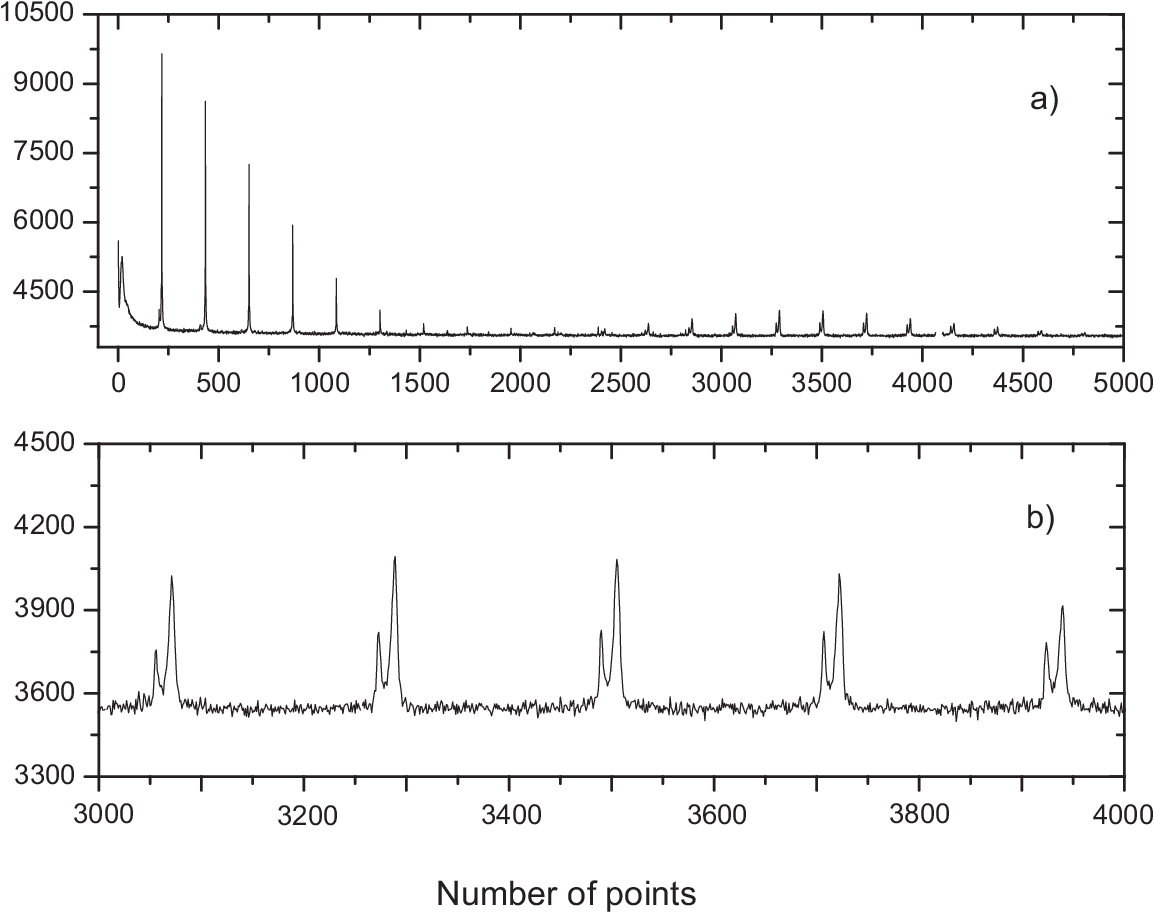}
	\caption{The obtained power spectrum of the pulsar PSR J0034-0721: $P_1$ = 0.943~s, $P_2$ = -63~ms, $P_3$ = 13.6$P_1$. On the top, it is a part of this power spectrum with satellite harmonics to the right of the main harmonics. The main harmonics are smaller than the modulation ones by about two times. The designation of the axes is the same as in Fig.\ref{fig:fig1}}
	\label{fig:fig7}
\end{figure}

In the summed up power spectrum of the pulsar  J0034-0721 (B0031-07) (Fig.\ref{fig:fig7}) both the modulation of the main harmonics (below) and the satellite harmonics (above) are clearly visible. The period $P_3$ is determined by the distance $r$ from the main harmonic: $P_3 = T / (r P_1$). On the obtained power spectrum in the range of point numbers up to 2,700, additional harmonics are not visible. The amplitude of the main ones becomes noticeable, and at the same time additional harmonics appear on the right at $n > 2,700$ the amplitude of which is about 2 times higher than the main harmonics. The center of the wide modulation envelope falls on n =3,255 which corresponds to $P_2 = -63$~ms. The satellite harmonics are located to the right of the main harmonic at a distance of $r = 16$ points which corresponds to the value of $P_3 = 13.6P_1$ and the drift occurs towards the beginning of the pulse, i.e. $P_2$ is negative. The analysis of the summed up power spectra shown in Fig.\ref{fig:fig7} gives the value $P_3 =13.6 \pm 0.4P_1$, corresponding to the drift A mode. Different drift
modes for this pulsar will be discussed in Section~\ref{sec_Result}. 

\begin{figure}
	\includegraphics[width=\columnwidth]{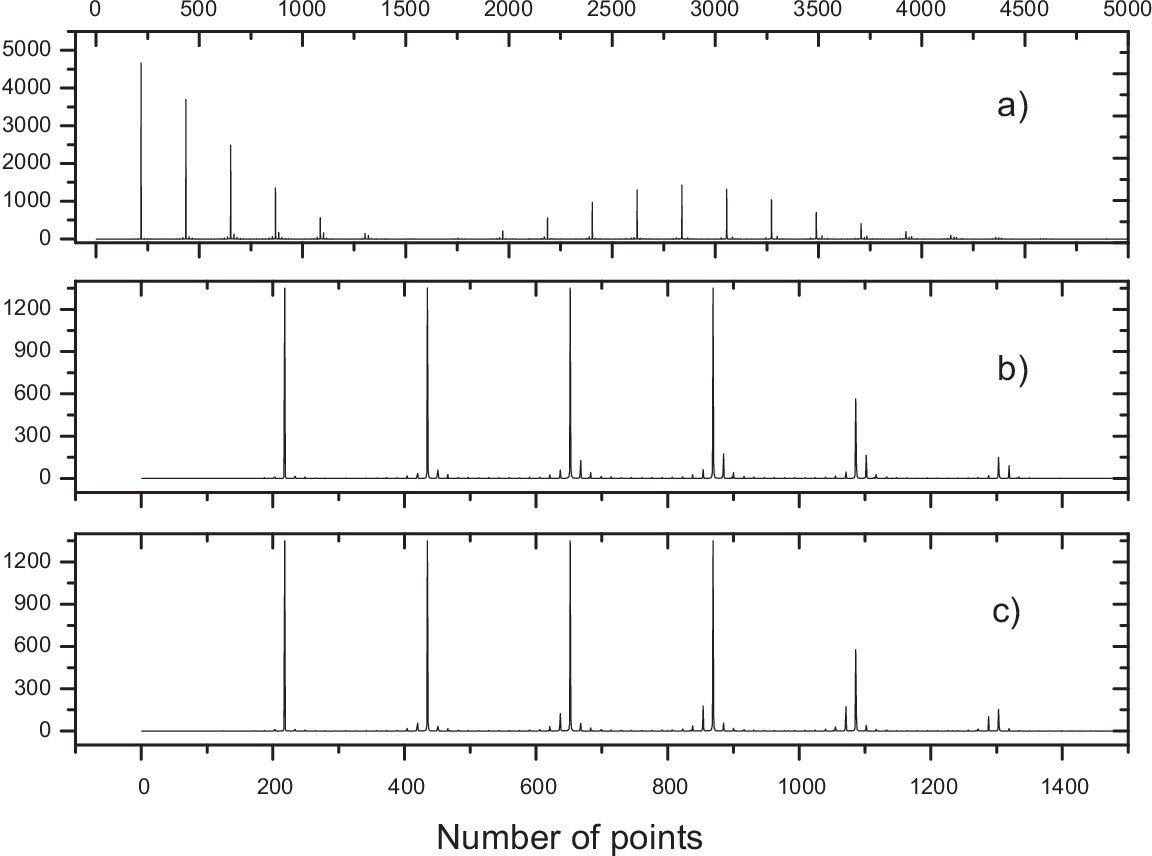}
	\caption{The model power spectra of the pulses drift with parameters of PSR J0034-0721 ($P_1 = 0.943$~s, $W_{0.5} = 12.5$~ms, $P_3 = 13.6{P_1}$): a) the model power spectrum in a wide range of harmonics c $P_2 = - 63$~ms; b) and c) the first harmonics of the power spectrum, increased in scale by 4 times, and with a different drift direction: b) $P_2 = -63$~ms, c) $P_2 = +63$~ms. The designation of the axes is the same as in Fig.\ref{fig:fig1}}
	\label{fig:fig8}
\end{figure}

The simulation shows that always when the direction of the drift of the pulses goes to the beginning of the profile ($P_2$  is negative) the satellite harmonic with higher amplitudes appears to the right of the main harmonic, and when the drift is towards the end of it ($P_2$ is positive) – to the left. Fig.\ref{fig:fig8} demonstrate this behavior of satellite harmonics. Here an example of models with parameters for J0034-0721 is shown for two cases: with a drift towards the beginning of the profile ($P_2 = -63$~ms (Fig.\ref{fig:fig8}b)) and towards the end of the profile ($P_2 = +63$~ms, Fig.\ref{fig:fig8}c), $P_3 = 13.6P_1$. Panel b in Fig.\ref{fig:fig8} corresponds to the summed up power spectrum. In Fig.\ref{fig:fig8}b,c, the first harmonics of the power spectrum are shown, increased in scale by 4 times, with the same scale for the horizontal axis. The shifted harmonics associated with the drift period $P_3$, have a significantly lower amplitude compared to the main harmonics at the beginning of the power spectrum. For $P_2 = -63$~ms, the satellite harmonic is located to the right of the main harmonic at a distance of 16 points, and for $P_2 = 63$~ms, it is to the left at the same distance as it should be when the drift is in different directions. In Fig.\ref{fig:fig8}a the power spectrum is shown in a wider range. For this power spectrum in the modulation ``hump'', only harmonics with offset by $r = 16$~points are visible, the amplitude of which exceeds the main harmonics by 10 times or more. 

Consequently, the most accurate definition of $P_3$ is  determination of the shifted harmonics position  to the right or left from the main ones associated with the period $P_1$. The position $r$ is determined with the precision of a discrete bin (up to one point), therefore, we will assume that it is no worse than $\pm 1/2$ of the discrete bin. Our each discrete bin is 0.004886 Hz ($1/T$). Accuracy of $P_2$ determination   is related to the accuracy of finding the centers of the ``humps'' of the slow modulation of the power spectra. We assume that the error in $P_2$ is determined with an accuracy of $\pm$ one harmonic from the center  position of the modulation envelope. The obtained parameters for $P_2$   and $P_3$ with their errors are given in Table\ref{tab:tab1}.

\section{Results and discussion}
\label{sec_Result}

In the paper \citetalias{Malofeev2018} where the new technique was proposed and tested, it was found that almost all the pulsars studied have modulation of the amplitudes of the main harmonics in the power spectra. This was interpreted as the presence of subpulses drifting and used to determine the value of $P_2$. The detection of satellites of harmonic peaks in power spectra   made it possible to measure the period $P_3$. The authors \citetalias{Malofeev2018} note that all new estimates both $P_2$ and $P_3$ need to be confirmed. Our simulation of pulses with a known pulsar period and the time interval $T$ has shown that determined period $P_2$ in \citetalias{Malofeev2018} from modulation of power spectrum harmonics is not associated with drift for 18 pulsars. The amplitude distribution of harmonics for them is explained by the corresponding value of $n=kT/P_1$. This effect was considered in Section~\ref{subsecP}. The list of these pulsars is  as follow: J0613+3721, J0826+2637, J0928+30, J1136+1551, J1635+2418, J1741+2758, J1758+3030, J1813+4013, J1907+4002, J1912+2525, J2018+2839, J2055+2209, J2113+2754, J2139+2242, J2208+4056, J2234+2114, J2305+3100, J2317+2149. In the paper \citetalias{Malofeev2018}, the sign of the drift period $P_2$ was not determined, and in this work, we present it for pulsars in which the values of $P_3$ are measured.

\begin{figure}
	\includegraphics[width=\columnwidth]{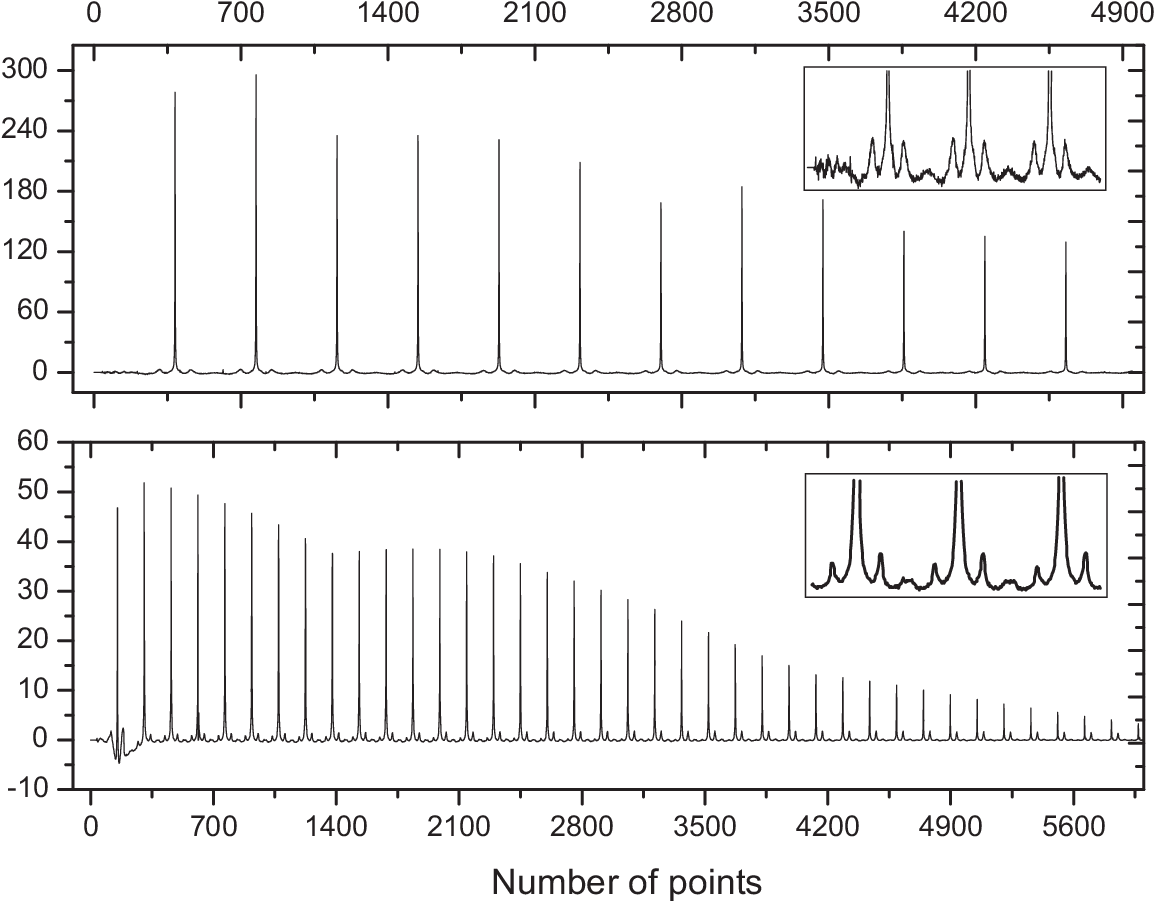}
	\caption{Summed up power spectrum of J0826+2637 (top) and J1921+2153 (bottom). The designation of the axes is the same as in Fig.\ref{fig:fig1}. Three harmonics of pulsars with magnification are shown in the upper right corner. The harmonic satellites are clearly visible in the enlarged pictures.}
	\label{fig:fig11}
\end{figure}

The presence of modulation in the power spectrum indicates either drift or the presence of subpulses with a distance between them which corresponds to the position of the maximum of this modulation. There may be cases when the drift may be present, but it is not regular or is very slow with $P_3 \gtrsim T / (2 P_1)$ and we cannot see shifted harmonics relative to the main ones ($r \lesssim 2$). Note that for J1921+2153, we got only the upper estimate of $P_2$ but we cannot exclude values of $P_2 \lesssim 40$~ms because in this case $k_{p_2} \gtrsim 5000$ and amplitude will be small. We got the value of $P_3=4.1P_1$ which correcpondings to the satellite harmonic shifted to the right of the main harmonics ($P_2$ is negative). The obtained power spectrum of PSR J1921+2153 is shown in Fig.\ref{fig:fig11}. PSR J0826+2637 has a satellites of a small equal amplitude (Fig.\ref{fig:fig11} top). On both sides of the main harmonics and we have defined $P_3$ indicated in Table\ref{tab:tab1}. There is no modulation corresponding to period $P_2$ (\citeauthor{Backer1970}, \citeyear{Backer1970}) found that this pulsar shows drifting in bursts, but the drift direction is different for different bursts. Our value of $P_3$ is in a good agreement with the value of \citetalias{Weltevrede2006}. The postcursor and interpulse known for this pulsar have small amplitudes: 30 and 65 times less than the amplitude of the main pulse at a frequency of 111 MHz (\citeauthor{Toropov2024}, \citeyear{Toropov2024}). In addition, the postcursor does not separate well from the main pulse with our time resolution of 12.5 ms. Therefore, these profile components do not affect the total power spectrum. 

The power spectrum of PSR J2234+2114 shows the satellites on both sides of the main harmonics about equal amplitude as for J0826+2637. It can be explained by the drift in different directions.  

$P_2$ value obtained for  J1136+1551 corresponds to the distance between the components of the average profile, and the period obtained in the works \citetalias{Weltevrede2006} and \citetalias{Malofeev2018} is not confirmed by us.

\begin{figure}
	\includegraphics[width=\columnwidth]{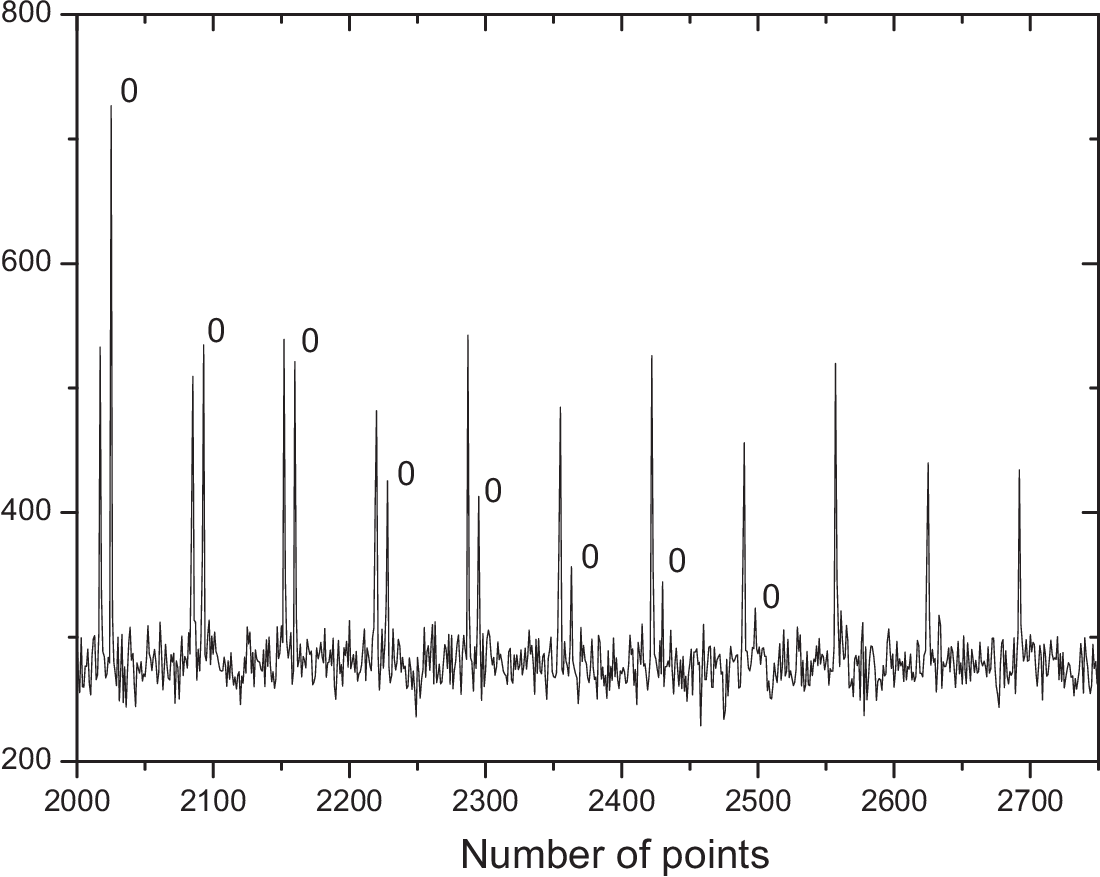}
	\caption{Part of the accumulated power spectrum of PSR J0323+3944. The ``0'' symbol indicates the main harmonic, in front of it – the harmonic shifted to the left, due to the drift towards the end of the profile. The designation of the axes is the same as in Fig.\ref{fig:fig1}}
	\label{fig:fig9}
\end{figure}

 Table\ref{tab:tab1} shows the estimates of the drift periods we have determined. If the pulsar falls into the classification \citetalias{Weltevrede2006}, it is indicated next to the designation of the pulsar in brackets. The first column of the table shows the names of pulsars. The asterisks next to the names indicates pulsars previously considered in the paper \citetalias{Malofeev2018}, for them the previously obtained values are given in brackets with the designation of \citetalias{Malofeev2018} in the corresponding columns. The second column gives a period. The third and fourth columns provide estimates of the drift period $P_2$ according to our analysis and the paper \citetalias{Weltevrede2006}. The fifth and sixth columns indicate the period $P_3$, determined in this paper and in \citetalias{Weltevrede2006}. If several periods $P_2$ and $P_3$ are defined in \citetalias{Weltevrede2006}, then they are also shown in the table. 

\begin{table*}
	\centering
	\caption{Pulsars with estimates of drift periods $P_2$ and $P_3$}
	\label{tab:example_table}
	\begin{tabular}{lcrrcc}
	\hline
	PSR	&	$P_1$, (s)	&	$P_2$ (LPA), (ms)	&	$P_2$ \citepalias{Weltevrede2006}, (ms)	&	$P_3$ (LPA)	&	$P_3$ \citepalias{Weltevrede2006}	\\
	\hline
J0034-0721 (Dif)	&	0.9429	&	$-63 \pm 5$ 	&	$-105^{+5}_{-131}$	&	$13.6 \pm 0.4$	&	$8.3 \pm 0.3$	\\
J0304+1932 (Dif*)	&	1.3875	&	$69.4^{+5}_{-6}$	&	$-135^{+27}_{-21}$	&		&	$5.2 \pm 0.3$	\\
J0323+3944*(Coh)	&	3.0321	&	$+118 \pm 10$ 	&	$152^{+42}_{-28}$	&	$8.44 \pm 0.05$ 	&	$8.4 \pm 0.1$	\\
					&			& ($112 \pm 13$ \citetalias{Malofeev2018}) & & ($8.4 \pm 1.3$ \citetalias{Malofeev2018}) & \\
J0528+2200* (Dif*)	&	3.7455	&	$-190 \pm 16$  	&	$-200^{+20}_{-90}$ 	&	   ($4.9 \pm 0.5$ \citetalias{Malofeev2018})	&	$3.8 \pm 0.7$	\\
					&			& ($200 \pm 10$ \citetalias{Malofeev2018}) & $520^{+580}_{-105}$ & & \\
J0611+3016*	&	1.4121	&	  	&		&	$3.6 \pm 0.5$ 	&		\\
			&			&		&		& ($2.5 \pm 0.2$ \citetalias{Malofeev2018}) & \\
J0826+2637* (Dif*)& 0.5307&  & $80^{+60}_{-10}$ & $5.2\pm 0.2$& $7\pm 2$ \\
                   &       &     &   & ($5.7\pm 0.4$ \citetalias{Malofeev2018})  \\
J1136+1551* (Dif*)	&	1.1879	&	$36 \pm 3$ 	&	$430^{+400}_{-50}$	&		&	$3 \pm 1$ 	\\
					&			& ($400 \pm 70$ \citetalias{Malofeev2018}) & & & \\
J1239+2452* (Dif*)	&	1.3824	&	$+62 \pm 5$ 	&	$61^{+4}_{-9}$  	&	$2.74 \pm 0.03$ 	&	$2.7 \pm 1$	\\
					&			& ($60 \pm 3$ \citetalias{Malofeev2018}) & $-77^{+5}_{-11}$ & ($2.7 \pm 0.1$ \citetalias{Malofeev2018}) & \\
J1532+2745  &      1.1248     &  $39\pm 3$      &       &       &       \\
            &                 & ($39\pm 2$ \citetalias{Malofeev2018}) &&& \\
J1721+3524	&	0.8219	&	$117^{+8}_{-11}$	&		&		&		\\
J1921+2153*(Dif)	&	1.3373	&	$< |-40|$	&	$-13 \pm 1$ 	&	$4.1 \pm 0.06$ 	&	$4.4 \pm 0.1$	\\
					&			&	($33\pm 1$ \citetalias{Malofeev2018})		& $-41 \pm 4$ & ($4.2 \pm 0.5$ \citetalias{Malofeev2018}) & \\
J2046+1540 (Dif*)	&	1.1383	&	$52 \pm 8$	&	$-22^{+1.6}_{-4.4}$	&		&	$18 \pm 6$	\\
J2227+3030    &  0.8424 &  $105\pm 10$ & & &  \\
              &         &  ($105\pm 15$ \citetalias{Malofeev2018})  &&& \\
J2234+2114 & 1.3574   &        &   & $25\pm 2$ & \\
           &          &        &   & ($23.3\pm 3.5$ \citetalias{Malofeev2018}) & \\
	\hline
 \label{tab:tab1}
\end{tabular}
\end{table*}

\begin{figure}
	\includegraphics[width=\columnwidth]{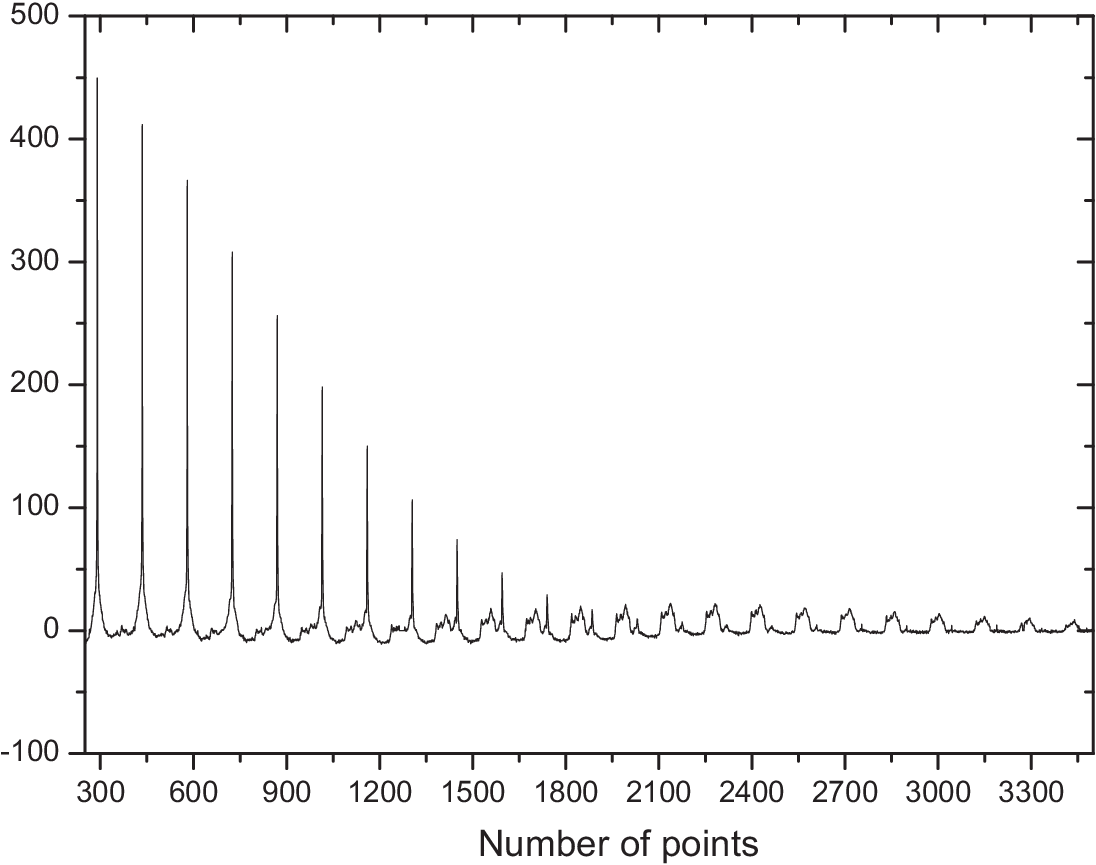}
	\caption{Summed up power spectrum of J0611+3016. The designation of the axes is the same as in Fig.\ref{fig:fig1}}
	\label{fig:fig10}
\end{figure}


As can be seen from Table\ref{tab:tab1}, for those sources for which there are no harmonic satellites in the power spectrum, we only give the value of $P_2$ without sign, while for pulsars with a positive value of $P_2$, we put the sign ``+'' (for case  we got the value for $P_3$). For J0304+1932, J1136+1551 and J2046+1540 in the paper \citetalias{Weltevrede2006} period $P_3$  was determined, whereas it could not be determined from observations on the LPA. These three pulsars belong to the class ``Dif*'' and, therefore, have unstable drift and wide details in the 2DFS spectrum.  For J1136+1551 and J0528+2250 measured values of $P_2$ correspond to the distance between the components at our frequency, as indicated above. Only one pulsar J0323+3944 from the table belongs to the class ``Coh'', and for it, the drift parameters are consistent within the error limits with the data in the paper \citetalias{Weltevrede2006}. Fig.\ref{fig:fig9} shows part of the summed up power spectrum for PSR J0323+3944. In this power spectrum, it is clearly seen that while the amplitude of the main harmonics is falling, the amplitude of the left-shifted harmonics associated with $P_3$ don't decrease and they become predominant. For 11 pulsars: J0034-0721, J0304+1932, J0323+3944, J0528+2200, J1136+1551, J1239+2452, J1532+2745, J1722+35, J1921+2153, J2046+1540, and J2227+3030 we have defined or redefined the drift period $P_2$. For 7 sources: J0034-0721, J0323+3944, J0611+3016, J0826+2637, J1239+2452, J1921+2153, and J2234+2114 drift direction and $P_3$ value  were determined.

For 9 pulsars from Table\ref{tab:tab1}, estimates of $P_2$  were previously obtained  in the paper \citetalias{Weltevrede2006}, and only for four of them (J0323+3944, J0528+2200, J1239+2452, and J1921+2153) our estimates coincide within the error limits. For J1921+2153 our upper limit for $P_2$ doesn't contradict with a value from \citetalias{Weltevrede2006}. J0034-0721 and J0304+1932 have a large errors for $P_2$ in \citetalias{Weltevrede2006} so it is difficult to compare its values. In the paper \citetalias{Weltevrede2006} seven of these 9 pulsars were identified as ``Dif'' or ``Dif*'', that is having no coherent drift. An unusual power spectrum is observed for J0611+3016. It can be seen from Fig.\ref{fig:fig10} that the harmonics due to the periodic emission of the pulsar have a low-level detail shifted to the left of the fundamental harmonic. The shift was defined as the difference from the maximum of it.    According to its shift, $P_3$ period was determined, listed in Table\ref{tab:tab1}. The period $P_2$ could not be determined because there is no expressed modulation in the power spectrum. There appears to be an irregular drift of subpulses towards the tail of the profile.

We used the simplest assumptions about subpulse drift to determine by simulated power spectra. In reality, drift behavior can be significantly more complicated. It is known that drift can be non-linear: two subpulses can have different velocities and direction, jumps in the phase of subpulses (\citeauthor{Edwards2003}, \citeyear{Edwards2003}), also the speed and direction of the drift can change over time. Pulsar J0034-0721 is known to have  3 drift modes A, B, C (\citeauthor{Huguenin1970}, \citeyear{Huguenin1970}), and the slope of the drift bands may vary from one band to another. In the paper (\citeauthor{Huguenin1970}, \citeyear{Huguenin1970}) parameters $P_2$ and $P_3$ have been determined  for these modes: 
$P_2 = -56\pm  6$~ms, $P_3 = 12.5\pm 0.5{P_1}$ (A mode); $P_2 = -58\pm 10$~ms, $P_3 = 6.8\pm 0.8P_1$ (B mode);  $P_2 = +56\pm  12$~ms, $P_3 = 4\pm  0.5P_1$~ms (C mode). The rarest was ``C'' mode. In the paper \citetalias{Weltevrede2006} it is noted that in ``A'' drift mode, pulsar has higher power ($P_3 = 12P_1$) and significantly lower power in ``B'' mode ($P_3 = 6P_1$), and there is no detail corresponding to ``C'' mode. Pulsar J0034-0721 has a mode-changing behavior and the range in $P_2$ quoted in \citetalias{Weltevrede2006} is most likely because of that.  The paper of the (\citeauthor{McSweeney2019}, \citeyear{McSweeney2019}) also indicates 3 drift modes, of which A mode has the same period $P_3 = 12.5\pm 0.8P_1$ as we have ($P_3 = 13.6 \pm 0.4P_1$). J0034-0721 has a nulling fraction of ~ 45\% (\citeauthor{McSweeney2019}, \citeyear{McSweeney2019}). Nullings can be for 50 or 150 pulsar periods and then for 400 $P_1$ they are not (Fig.~3 from (\citeauthor{McSweeney2019} (\citeyear{McSweeney2019})). Our one session has a duration of 3.4 minutes and different nulling duration will be implemented on different days. With the accumulation of power spectra, the days with the smallest fraction of nulling during the observation will make the greatest contribution to the total power spectrum. Simulation of pulses with obtained drift parameters for J0034-0721 ($P_2$ = -63 ms and $P_3$ = 13.6$P_1$) and a nulling of 45\% showed that nulling does not affect the analysis of our power spectrum (see Fig.\ref{fig:fig12}). It is clear that our simple model does not describe the complex behavior of drifting subpulses, but it explains the resulting summed up power spectrum quite well, correctly determines the drift direction and position of the harmonic satellites relative to the main harmonics. The power spectrum accumulated over a long period of time reflects the most frequently realized drift parameters. In this case, the obtained drift parameters are consistent with the drift mode ``A'', which means that it is most often manifested for J0034-0721. 

For J0304+1932 in the paper \citetalias{Weltevrede2006} there is a wide drift detail in 2DFS for the tail component of the profile till the boundary of the spectrum ($2 P_1$, Nyquist frequency), but there is no drift in the main component. For J1136+1551, period $P_3$ also has a value close to $2 P_1$, and as for J0304+1932, $P_2$, determined in \citetalias{Weltevrede2006}, have large values of $P_2$ which 
most likely indicate amplitude modulation, not phase modulation as it was pointed in \citetalias{Weltevrede2006}. Nonlinear drift or random changes in the direction of drift during observation can also lead to large values of $P_2$. 
As noted in \citetalias{Weltevrede2006}, J2046+1540 has a wide drift detail for the tail component, as well as the value of $P_3$  may vary relative to the average value during observations. In our paper $P_3$ is not determined. 

For four sources: J0928+3037, J1635+2418, J2018+2839, and J2305+3100, observed in \citetalias{Malofeev2018}, period $P_3$ is not confirmed in our analysis. J0928+3037 and J1635+2418 were not reported in \citetalias{Weltevrede2006}. J2018+2839 has 2 different drift modes in \citetalias{Weltevrede2006}. For PSR J2305+3100, the detail in the power spectrum (\citetalias{Weltevrede2006}) splits in the vicinity of the Nyquist frequency: $P_3 = 2P_1$, and this is interpreted as the presence of two directions of drift. 
The direction of drift changes during observation. Out of the 41 pulsars in our list, 9 have measurements of $P_3$ in \citetalias{Weltevrede2006}. We did not detect $P_3$ for three of them. All these sources belong to the class ``Dif*'' and have a complex pattern of drift. For 5 pulsars from our Table\ref{tab:tab1} we have a good agreement $P_3$ values with \citetalias{Weltevrede2006} and (\citeauthor{Huguenin1970} (\citeyear{Huguenin1970}). The literature provides an analysis of the drift behavior of pulsars based on single observations over a period of several hours at best. Based on such observations, it is difficult to conclude what part of the time the pulsar is in one or another drift mode for a long time. As a rule, our total power spectrum reflects the most frequently realized drift mode over an interval of about 5 years (total observation time). 

\section{Conclusions}

For the search of drift periods, the summed up power spectra of known pulsars, obtained by us as a result of processing monitoring data, were considered. Our work is devoted to the study of the drift behavior of pulsar pulses. It is a logical development of the method of using the summed up power spectra of pulsars to measure the drift periods of subpulses \citepalias{Malofeev2018}. Our analysis showed that drift behavior was not detected for most of the pulsars studied.  The distribution of the amplitudes of the main harmonics in many cases does not differ from their distribution without the presence of drift. The summed up power spectra of pulsars in the presence of drift are also well described by the proposed model. 
In the later case, the drift is manifested by the presence of modulation of the main harmonics in the power spectrum, with the  frequency inverse to $P_2$. The period $P_3$  manifests itself in the summed up power spectra as additional harmonics shifted from the main harmonics in one direction or the other. The amplitude of these harmonics may exceed the amplitude of the main ones. The presence of these harmonics on the left means that the subpulses drift to the beginning of the profile ($P_2$ is negative). A shift to the right means that the drift occurs towards the end of the profile($P_2$ is positive). 

The drift parameters obtained from the accumulated power spectra correspond to the drift behavior most often realized over a long period of observations. Apparently, there are few pulsars with regular coherent drift. This follows from the paper  \citetalias{Weltevrede2006}, in which out of 42 pulsars with drift behavior, only 19 show narrow details in two-dimensional spectra (class ``Coh'' – coherent), moreover, 7 of them have 2 values of $P_2$ significantly different from each other, but having the same $P_3$. Accordingly, the rate of displacement of subpulses changes quite sharply over time. In the power spectra accumulated over many days, such drift behavior, of course, will not give a clear picture. Our analysis showed that the model power spectra of pulsars without drift describe well the summed up power spectra obtained for 18 pulsars in \citetalias{Malofeev2018} and, thereby, significantly reduced the number of sources with a confirmed period $P_2$ that was reported by \citetalias{Malofeev2018}. 
Using the considered technique, it was possible to determine the period $P_2$ for 11 out of 68 pulsars  were studied on LPA LPI. The drift direction and $P_3$ value we got for 7 pulsars. Among 13 sources common with paper  \citetalias{Weltevrede2006}, for five sources $P_2$ is not determined, for five pulsars the values coincide with the previously reported measurements, and for three pulsars they do not coincide. The period $P_3$ could be measured for 9 pulsars. Among these out of 8 sources common with the work of \citetalias{Weltevrede2006}, the period  coincides for 5 sources, and for three it is not determined. 

\section*{Acknowledgements}
We are grateful to the anonymous referee for a thorough reading of the manuscript and for the comments that made it possible to improve our work. The study was carried out at the expense of the Russian Science Foundation grant 22-12-00236, https:// rscf.ru/ project/ 22- 12- 00236/.

\section*{Data availability}
The PUMPS survey is not finished yet. The raw data underlying this paper will be shared on reasonable request to the corresponding author.

\newpage
\section{Appendix}
\appendix
List of 41 investigated pulsars \label{Appendix A}.

J0034-0721; J0048+3412; J0051+0423; J0146+31; J0220+3626; J0304+1932; J0528+2200; J0608+1635; J0629+2415; J0659+1414; J0837+0610; J0922+0638; J0946+0951; J0953+0755; J1238+2152; J1242+39; J1313+0931; J1404+1159; J1543+0929; J1543-0620; J1614+0737; J1627+1419; J1645-0317; J1645+1012; J1721+3524; J1740+1311; J1821+4147; J1823+0550; J1844+1454; J1920+2650; J1931-0144; J1932+1059; J1946+1805; J1952+1410; J2007+0910; J2046+1540; J2116+1414; J2212+2933; J2215+1538; J2219+4754; J2253+1516.

\end{document}